\documentstyle[11pt]{article}
\input{psfig} 
\begin{document}
\begin{center}
{\Large\bf THE NUCLEAR STAR CLUSTER OF THE MILKY WAY: STAR FORMATION, DYNAMICS 
AND 
CENTRAL BLACK HOLE}

\vspace{\baselineskip}
Reinhard Genzel\\
Max--Planck--Institut f\"ur extraterrestrische Physik\\
Garching, FRG\\
and\\
Department of Physics\\
University of California at Berkeley, USA
\end{center}

\begin{abstract}
High spatial resolution, near-infrared imaging and spectroscopy of the nuclear 
star cluster have given key new insights about the dynamics, evolution and mass
distribution in the Milky Way Center. The central parsec is powered by a cluster
of hot, massive stars which must have formed a few million years ago. Either 
star formation was triggered in the central parsec by the infall of a very dense
cloud, or a dense, young star cluster formed outside of the central parsec sank
rapidly into the nuclear region through dynamical friction. The presence of 
luminous asymptotic giant branch (AGB) stars suggests that there were earlier 
such star formation episodes.

Measurements of radial and proper motions for more than 200~stars delineate the
stellar dynamics to a scale of a few light days from the dynamic center which is
coincident with the compact radio source (SgrA$^*$) within 
$\sim$0.1$^{\prime\prime}$ (800\,AU)\@. The stellar velocities increase toward 
SgrA$^*$ with a Kepler law (to $>$1000\,km/s for the innermost stars), 
implying the presence of a three million solar mass central (dark) mass. The 
observations make a compelling case that this mass concentration is a black hole
which is currently accreting at a low rate or radiating at low efficiency. With
the exception of the young, massive stars the velocity field of the central 
stellar cluster is close to isotropic. The young stars are characterized by a 
turbulent rotation pattern that still carries the imprint of the angular 
momentum distribution in the original cloud or star cluster.
\end{abstract}

\section{INTRODUCTION}
The nucleus of the Milky Way (distance $\sim$8\,kpc) is one hundred times closer
to us than the nearest large external galaxy and a thousand times closer than 
the nearest active galactic nuclei. We can study physical processes in our own 
Galactic Center at a level of detail that will never be reached in the more 
distant, but often also more spectacular systems. What powers these nuclei and 
how do they evolve? What are the properties of the star clusters located in 
their cores? Is star formation happening there? Do massive black holes reside at
the dynamical centers and how do they form and evolve? In the present paper we 
summarize what is presently known in the Galactic Center about some of these key
issues. For more detailed recent reviews see Genzel, Townes and Hollenbach 
(1994), Morris and Serabyn (1996), Mezger, Duschl and Zylka (1996) and Genzel 
and Eckart (1998). 

\section{THE NUCLEAR STAR CLUSTER: PROPERTIES AND\\EVOLUTION}

The total UV and visible luminosity of the central parsec is 30 to 50~million 
solar luminosities (Telesco et al.~1996, Mezger et al.~1996). About ten percent
of that luminosity is emitted in the hydrogen ionizing continuum ($>$13.6\,eV),
with a characteristic temperature of about 30,000 to 35,000\,K (Lacy et 
al.~1980, Shields and Ferland 1994, Lutz et al.~1996). What powers this fairly 
low excitation, ionized (H{\small II}) region and what are the properties of the
central star cluster? During the past decade high resolution near-infrared 
observations have significantly improved our knowledge of the distribution and 
characteristics of the nuclear stellar cluster and provided a fairly unambiguous
picture of the energetics and evolution of the region. Through the advent of 
sensitive, large format infrared detector arrays and speckle/adaptive optics 
imaging it has become possible to image the central parsec at diffraction 
limited resolution. With $\mathrm{D}=4$ to 10m diameter telescopes the FWHM 
resolution that can be attained is 
$\sim0.13^{\prime\prime}(3.5\mathrm{m}/\mathrm{D})$
or 0.005\,pc at 2.2\,micrometers (K-band) (Eckart et al.~1992, 1993, 1995, 
Genzel et al.~1997, Davidge et al.~1997, Ghez et al.~1998, 2000). Rieke (1999) 
and Stolovy et al.~(1999) describe the first HST-NICMOS results. The best 
current images resolve the near-infrared emission of the central parsec into 
almost 1000~stars with K-band magnitudes $\mathrm{m}(\mathrm{K})<15$ to 16 
(Fig.~1). Thus
all red and most blue supergiants, all red giants (including bright, asymptotic
giant branch (AGB) stars) of spectral type later than K5, and all main sequence
stars earlier than about B2 should be visible in Fig.~1. Nevertheless this image
still only samples about 0.1\% of the total stellar content of the cluster. 
Further progress can be expected from deeper near-infrared images that will be 
obtainable in the next few years with the adaptive optics systems on the Keck, 
Gemini and VLT. 

The K-band surface brightness and surface density increase approximately with 
the inverse of the distance to about 1$^{\prime\prime}$ from the central compact
radio source SgrA$^*$ (see below). The centroid of the stellar surface density 
distribution is within 0.1$^{\prime\prime}$ of SgrA$^*$ (Ghez et al.~1998). 
Near SgrA$^*$ the most prominent feature is a group of two dozen or more bright
stars (the `IRS16' complex) centered 1--2$^{\prime\prime}$ east of the radio 
source. About 3.5$^{\prime\prime}$ south-west of SgrA$^*$ lies another compact 
group of bright stars (the `IRS13' complex). There is also an additional 
enhancement of fainter stars within $<$1$^{\prime\prime}$ of SgrA$^*$ (Eckart 
et al.~1995, Ghez et al.~1998). This so called `SgrA$^*$ cluster' (Fig.~1) is 
particularly striking on the 0.05$^{\prime\prime}$ resolution K-band image 
taken by Ghez et al.~(1998) with the 10m Keck telescope (right inset of Fig.~1).
The core radius (=\,radius at which the surface density is half of the central 
value) of the $\mathrm{m}(\mathrm{K})<15$ stellar surface number density 
distribution is $\sim$2$^{\prime\prime}$ (0.08\,pc: Genzel et al.~2000, 
Alexander 2000). It is debatable whether this value is also a good estimate of 
the core radius of the overall (old) stellar cluster since the observed stars 
only sample the brighter (and more massive) members of the cluster, as discussed
above. Allen (1994) and Rieke and Rieke (1994) have deduced larger values of the
core radius (0.5 to 0.8\,pc) from the near-IR surface brightness distribution of
the late type stars only. Genzel et al.~(1996) have proposed a core radius of 
$\sim$0.4\,pc as a compromise between these extremes in which case the stellar 
density in the core is about $4\times10^6\,\mathrm{M}_\odot/\mathrm{pc}^3$. 
There are a total of about $10^{5.5}$ stars within the core. The SgrA$^*$
cluster may represent a central stellar `cusp' associated with the radio source.
From an analysis of recent high resolution, near-IR data sets Alexander (2000, 
and references therein) concludes that a cusped distribution with a power law 
density distribution of exponent $-$1.5 to $-$1.75 is a better fit (at 
$>$2$\sigma$) to the data than a distribution that has a flat central core. If 
this conclusion is correct, the stellar density in the SgrA$^*$ cluster may be 
$10^8\,\mathrm{M}_\odot/\mathrm{pc}^3$ or greater, compared to the 100~times 
lower value averaged over the central 0.4\,pc. One consequence then would be 
that the absence of bright late type giants in the innermost few arcseconds 
(Sellgren et al.~1990, Genzel et al.~1996, Haller et al.~1996) may be explained
by envelope destruction in close impact, giant-dwarf or giant-binary collisions
(Alexander 2000, Davies et al.~1998). Another important consequence is that with
future sensitive adaptive optics and interferometric imaging 
($\mathrm{m}(\mathrm{K})\sim 19$) it should be possible to detect up to 
100~stars
residing within 0.1$^{\prime\prime}$ and several stars within 
0.01$^{\prime\prime}$ of SgrA$^*$.

Another important aspect has been the discovery of a cluster of about 25~bright
He{\small I}/H{\small I}-emission line stars centered on the IRS16/IRS13 complex
(Forrest et al.~1987, Allen et al.~1990, Krabbe et al.~1991). As shown in 
Fig.~2 several of the brightest members of the IRS16 complex are 
He{\small I}-stars, as is IRS13E (Krabbe et al.~1995, Eckart et al.~1995, 
Libonate et al.~1995, Blum et al.~1995b, Genzel et al.~1996, 2000, Tamblyn et 
al.~1996, Morris et al.~2000). From non-local thermodynamic equilibrium (NLTE),
stellar atmosphere modeling of the observed emission characteristics Najarro et
al.~(1994, 1997, 1999) have inferred that the He{\small I}-stars are moderately hot (17,000 to 30,000\,K) and very luminous (1 to 
$30\times 10^5\,\mathrm{L}_\odot$). Their helium rich surface layers are 
expanding as powerful stellar winds with velocities of 200 to 800\,km/s and 
mass loss rates of 1 to $70\times 10^{-5}\,\mathrm{M}_\odot/\mathrm{year}$. 
Fig.~3 shows the location of these stars in a Hertzsprung-Russell diagram, 
along with stellar evolutionary tracks (Meynet et al.~1994) for twice solar 
metallicity element abundances probably appropriate for the Galactic Center: 
Lacy et al.~1980, Shields and Ferland 1994, but see Carr et al.~2000, 
Ramirez et al.~2000 and below). The He{\small I}-stars thus appear to be blue 
supergiant stars of initial mass 40 to $>$100\,M$_\odot$ that have evolved off 
the main sequence. Figure 2 shows that nucleosynthesis products (He, N, C) are 
clearly present in their outer atmospheres. They are probably on their way to 
becoming hot Wolf-Rayet stars and then to exploding as supernovae. Empirically 
they are similar to late WN/WC~stars, Luminous Blue Variables and Of(pe) 
supergiants (Allen et al.~1990, Krabbe et al.~1991, Najarro et al.~1994, 
Libonate et al.~1995, Blum et al.~1995a,b, Tamblyn et al.~1996). Combining the 
contributions from all its members, the He{\small I}-star cluster can plausibly
account for most of the bolometric and Lyman-continuum luminosities of the 
central parsec (Krabbe et al.~1995, Najarro et al.~1997). As yet unobserved 
hotter Wolf-Rayet and O~stars are required, however, to account for the helium 
ionizing luminosity of SgrA West. The He{\small I}-star cluster also provides 
in excess of $10^{38}\,$erg/s in mechanical wind luminosity which probably has 
a significant impact on the gas dynamics in the central parsec (Genzel, 
Hollenbach and Townes 1994). Krabbe et al.~(1995) have fitted the properties of
the massive early type stars in the central parsec by a model of a star 
formation `burst' between 2 and 9~million years ago in which a few hundred 
OB~stars and perhaps a few thousand stars in total were formed. This conclusion
is in excellent agreement with earlier proposals by Rieke and Lebofsky (1982), 
Lacy, Townes and Hollenbach (1982) and Allen and Sanders (1986). In the 
analysis of Krabbe et al.~the He{\small I}-stars are the most massive cluster 
members that have already evolved off the main sequence. In this scenario the 
central parsec is now in the late, wind-dominated phase of the burst. A similar
object is the R136 star cluster powering the 30~Doradus nebula in the Large 
Magellanic Cloud. The presence of a number of highly dust-enshrouded and 
spatially resolved infrared sources (e.g.~IRS1, 3 and 21, apparent in the upper
right of Fig.~3, Krabbe et al.~1995, Ott et al.~1999, Tanner et al.~1999) may 
indicate that stars are still forming at the present time. However, the star 
formation activity appears to be significantly less now than during the peak of
the burst. Likewise the small number of red supergiants (1 to 3 in the central 
parsec, Blum et al.~1996a) shows that the star formation rate prior to 10 or 
more million years ago also was substantially smaller. The relatively large 
number ($\sim$30 in central parsec, Genzel et al.~1996) of very cool 
($<$3000\,K, Blum et al.~1996a) and very bright red giants with luminosities 
$10^3$ to $10^4$\,L$_\odot$ (=\,AGB stars, apparent in Fig.~3 as a group to the
right from the top of the giant branch) may signify other such starburst 
episodes that probably  happened between 100 and 1000~million years ago 
(Haller and Rieke 1989, Krabbe et al.~1995, Blum et al.~1996b, Sjouwerman et 
al.~1999). 

The present gas density in the central parsec is far too low for gravitational 
collapse of gas clouds to stars in the presence of the strong tidal forces 
(Morris 1993). Perhaps the most recent episode of star formation was triggered 
by infall of a particularly dense gas cloud about 10~million years ago. This 
cloud may then in addition have been compressed by shocks and cloud-cloud 
collisions in the central parsec, thus triggering gravitational collapse. The 
cloud-infall model is also supported by an overall counter-rotation (in the 
sense of Galactic rotation) of the He{\small I}-star cluster 
(Genzel et al.~1996, 2000 and below). Another possibility is that a young star 
cluster came into the nuclear environment on a highly elliptical or parabolic 
orbit. If that star cluster was initially dense enough to have been tidally 
stable, it could have rapidly sunk in due to dynamical friction, followed by 
tidal disruption in the innermost region (Gerhard 2000). As an alternative to 
the starburst scenario Eckart et al.~(1993) had considered formation of massive
stars by sequential merging in star-star collisions. Fokker-Planck modeling of 
an evolving Galactic Center type, dense cluster shows, however, that merging 
can account for only $\sim$10 20\,M$_\odot$ stars and no $>$30\,M$_\odot$ stars
(Lee 1994). The basic reason is that in the calculations a sufficiently dense 
stellar core (density $10^7\,\mathrm{M}_\odot/\mathrm{pc}^3$ or greater) cannot
be maintained for a long enough time to build up very many massive stars. 
Further Morris (1993) had suggested that the He{\small I}-stars are not 
classical blue supergiants at all but transitory objects that have been created
in collisions between ($\sim$10\,M$_\odot$) stellar black holes and solar mass,
red giants. Both accounts of the He{\small I}-stars just cited are very 
specific to the high density environment of the central parsec. However, the 
`Quintuplet' and `Arches' clusters several tens of parsecs north of SgrA have a
massive star content and evolutionary state (including He{\small I}-stars) 
remarkably similar to that of the central, high density nuclear region (Cotera 
et al.~1996, Figer et al.~1995, 1999a,b). Further Ott et al.~(1999) found that 
the He{\small I}-star IRS16SW is an eclipsing binary with a minimum mass of 
100~solar masses. These facts and the presence of heavy element nucleosynthesis
products discussed above (Fig.~2) strongly favor the star formation model over 
the other scenarios. If the Galactic Center is representative of other galactic
nuclei as well, nuclear star formation may be a dynamic and highly time 
variable process that is the result of a complex interplay of the triggering 
effects of cloud infall and compression on the one hand, and of the destructive
effects of stellar winds and supernova explosions on the other hand. 

Figure~4 shows a K-band spectrum of the central 0.6$^{\prime\prime}$ centered 
on SgrA$^*$, obtained with the ISAAC infrared spectrometer on the ESO-VLT (from
Eckart et al.~1999, see also Genzel et al.~1997, Figer et al.~2000). The 
integrated spectrum of the SgrA$^*$ cluster is blue and featureless and 
requires stars hotter than K-type giants. Given the typical K-band magnitudes 
$(\mathrm{m}(\mathrm{K})\sim 14$--16) the SgrA$^*$ cluster members thus are 
likely early B or late O~stars (Genzel et al.~1997). 

An interesting new application of infrared spectroscopy is the study of element
abundances in the Galactic center stars. As mentioned above, the metallicity of
the SgrA West H{\small II} region is probably greater than solar, with a best 
estimate of twice solar abundances (e.g.~Shields and Ferland 1994). Najarro et 
al.~(1999) have reported that the Mg- and Fe-abundance in the luminous `Pistol'
blue supergiant in the Quintuplet cluster is also at least twice solar. Recent 
high resolution near-IR spectroscopy of IRS7 (Fig.~5, Carr et al.~2000, Ramirez
et al.~2000) and half a dozen other M-supergiants within 30\,pc of SgrA$^*$, on
the other hand, results in a near solar, Fe-abundance. It is not clear yet 
whether these results are in conflict with each other, or whether the 
uncertainties in the analysis allow a common solution.  	

In summary, the stellar and nebular observations of the central parsec are quite
well described by a modest, aging starburst that currently dominates the 
energetics of the SgrA West H{\small II} region. Several puzzles remain. 
OB main sequence stars and early Wolf-Rayet stars have not yet been 
unambiguously observed. It is surprising that rare, short-lived stars (such as 
WN9/Ofpe and LBVs) dominate the stellar census (e.g.~Tamblyn et al.~1996). 
Quantitative estimates of the emerging (E)UV spectral energy distribution of an
evolving star cluster based on current tracks predict an overall increase of 
the effective temperature of the radiation field a few million years after the 
burst (Lutz 1999, see Fig.~3). This is due to hot WN/WC~stars dominating the 
EUV radiation field that controls the excitation of the H{\small II} region. 
This prediction is at odds with the observations; they indicate that most of 
the energetics/excitation of the SgrA H{\small II} region can be plausibly 
accounted for by the He{\small I} emission line cluster. It appears that the 
evolutionary tracks do not properly describe the stellar census in the central 
parsec. One issue in this context is the role of stellar rotation in the 
appearance and characteristics of the massive stars (e.g.~Langer and Maeder 
1995, Heger and Langer 2000). Rotational mixing may bring up efficiently 
nucleosynthesis products into the outer atmospheres of the stars. Hanson et 
al.~(1996) have proposed that the enhanced nitrogen abundances apparent in the 
spectra of several of the Galactic center stars (e.g.~IRS13E in Fig.~2, similar
to ON supergiants) may be related to this effect. 

\section{THE COMPACT RADIO SOURCE SGRA$^*$}  
Ever since its original discovery the compact, nonthermal radio source 
SgrA$^*$ at the core of the nuclear star cluster has been the primary candidate
for a possible massive black hole at the Galactic Center, in analogy to compact
nuclear radio sources in other nearby normal galaxies (Lynden-Bell and Rees 
1971). In fact ever more detailed radio measurements have confirmed the unique 
nature of SgrA$^*$ in the Galaxy. Recent very long baseline radio 
interferometry (VLBI) observations in the mm-range show its intrinsic size, 
after correction for interstellar scattering, to be less than about 3\,AU 
(Bower and Backer 1998, Lo et al.~1999, Krichbaum et al.~1999).  Yet SgrA$^*$ 
is relatively faint in any wavelength range other than the cm-mm band. Using 
several several bright stars with radio masers in their envelopes (present on 
both radio and near-infrared maps) Menten et al.~(1997) have been able to 
register SgrA$^*$ on near-infrared maps with an uncertainty of 
$\pm$30\,milli-arsec (asterisk in Fig.~1). SgrA$^*$ is located near the 
centroid of the SgrA$^*$ cluster but it is not coincident with any steady 
source of $\mathrm{m}(\mathrm{K})<16$ (Genzel et al.~1997, Ghez et al.~1998). 
On the June 1996 and July 1997 NTT images (Genzel et al.~1997) there is a 
$\mathrm{m}(\mathrm{K})\sim 15.5$ source at the nominal position of SgrA$^*$, 
possibly implying a time variable source associated with SgrA$^*$. 
Alternatively one of the nearby faint stars may have moved there (Ghez et 
al.~1998). Nevertheless it is fairly clear that SgrA$^*$ has been 
infrared-quiet during the past one or two decades. This limits its infrared 
luminosity to less than a few thousand solar luminosities. Recent 
arcsecond-resolution observations with CHANDRA have established that there is a
(weak), compact keV-source at the position of SgrA$^*$ (Baganoff et al.~2000). 
Its luminosity in the 1--10\,keV band is less than a few L$_\odot$. 
Observations with ASCA and GRANAT suggest that SgrA$^*$'s X-ray luminosity may 
have been larger in the past few hundred years (a few $10^5\,$L$_\odot$, 
Koyama et al.~1996) but still orders of magnitude smaller than the Eddington 
rate of a million solar mass black hole (Sunyaev et al.~1993).

If SgrA$^*$ is a black hole, it must be radiating at a surprisingly low level.

\section{GAS AND STELLAR DYNAMICS: EVIDENCE FOR CENTRAL DARK MASS}
The evidence for a (dark) central mass in the Galactic Center thus is based 
entirely on the gas and stellar dynamics. While the velocities of gas clouds 
and of stars are approximately constant outside of a few parsec --- as expected
if the mass is dominated by the dense, near-isothermal nuclear star 
cluster --- velocities are observed to increase with a Kepler law within the 
inner core (e.g.~Genzel and Townes 1987). The first evidence for this increase 
in gas velocities came from mid-infrared spectroscopy of the 12.8\,micrometer 
[Ne{\small II}] emission line by Wollman et al.~(1977) and Lacy et al.~(1980). 
These authors and others following interpreted the $>$250\,km/s gas velocities 
as signaling a concentration of non-stellar mass in the Galactic Center, 
possibly caused by a few million solar mass black hole at the dynamic center 
(Lacy et al.~1982, Serabyn and Lacy 1985). However, gas motions can be affected
by magnetic, frictional and wind forces, in addition to gravity. Stellar 
velocities are required for an unambiguous determination of the mass 
distribution. Beginning with the pioneering work of Rieke and Rieke (1988), 
McGinn et al.~(1989) and Sellgren et al.~(1990) ever better stellar velocities 
from Doppler shifts of stellar absorption and emission lines have become 
available during the past decade, fully supporting the earlier measurements of 
gas velocities and very substantially strengthening the evidence for a compact 
central dark mass in the Galactic center (Rieke and Rieke 1988, McGinn et 
al.~1989, Sellgren et al.~1990, Lindqvist et al.~1992, Krabbe et al.~1995, 
Haller et al.~1996, Genzel et al.~1996). The most recent determinations by 
Krabbe et al.~(1995), Haller et al.~(1996) and Genzel et al.~(1996) are all in 
excellent agreement and show a significant increase of stellar radial velocity 
dispersion from about 55\,km/s at 5\,pc to about 180\,km/s at 0.15\,pc. 

A breakthrough in the evidence for a central dark mass occurred when the first 
measurements of stellar proper motions became available. These sample the 
stellar dynamics to a few light days from SgrA$^*$. Eckart and Genzel (1996, 
1997) and Genzel et al.~(1997) reported proper motions for more than 50~stars 
between $\sim$5$^{\prime\prime}$ (0.2\,pc) and $\sim$0.1$^{\prime\prime}$ 
(0.004\,pc) from SgrA$^*$ (Eckart and Genzel 1996, 1997, Genzel et al.~1997). 
The MPE group originally derived their results from 0.15$^{\prime\prime}$ 
speckle images obtained on the ESO New Technology Telescope (NTT) in 8~observing
runs at least once a year between 1992 and 1997. Independently, Ghez et 
al.~(1998) reported proper motions for 90~stars between 0.1$^{\prime\prime}$ and
4.3$^{\prime\prime}$ from SgrA$^*$. The UCLA group determined their results 
from 0.05$^{\prime\prime}$ resolution speckle imaging with the 10m Keck 
telescope in three epochs between 1995 and 1997. More recently, Eckart et 
al.~(1999) and Genzel et al.~(2000) have updated their proper motion data set 
(including two more NTT runs in 1998 and 1999 and combining the NTT data with 
the Ghez et al.~(1998) Keck data) to yield more than 100~proper motions of 
significantly improved quality (Fig.~6). Likewise Ghez et al.~(2000) have also 
updated and improved their proper motions, including the first detection of 
curvature in the proper motion trajectories of three stars very close to 
SgrA$^*$. With a few exceptions the proper motions deduced independently by the
two groups are in excellent agreement. 

\section{CONSTRAINTS ON ANISOTROPY}
For those 32~stars between 1 and 5$^{\prime\prime}$ from SgrA$^*$ for which 
both radial and proper motions are available, the deduced velocity dispersions 
in the three spatial directions agree very well (Genzel et al.~2000, for an 
adopted 8\,kpc distance). Moreover and more significantly, for all 104~stars 
with proper motions in the list of Genzel et al.~(2000), the sky-projected, 
tangential and radial velocities of each star are also in good agreement with 
an isotropic distribution (Fig.~7, lower right). Large scale anisotropy of the 
velocity field does not play a major role in the central parsec of the Galaxy. 
This fact substantially increases the robustness of the mass distribution 
discussed in the next section.

The picture changes if only the proper motions of the early type, massive stars
are considered. 11 out of 12~He{\small I} emission line stars within 
5$^{\prime\prime}$ from SgrA$^*$ are on projected tangential orbits (Fig.~7, 
upper left). Most of the He{\small I} emission line stars and the brighter 
members of the IRS16 complex follow a clockwise (on the sky) and 
counter-rotating (with respect to the Galaxy) coherent streaming pattern 
(Fig.~6). This pattern indicates that the young stellar component is in overall
rotation, albeit with large, local random motions. The rotation is likely a 
remnant of the original angular momentum distribution of the cloud (or young 
star cluster) the massive stars came from. The age of the massive stars is 
significantly less than the relaxation time at 0.5\,pc (10 to 30~million years).

For the late type stars the proper motions are fully consistent with isotropy 
(Fig.~7, lower left), as expected for their greater age. Still, there are some 
indications in the radial velocity data (Ott et al.~2000) for bright late 
type (=\,AGB) stars projected in a similar part of the sky to also have similar
radial velocities. This clumping in phase space is puzzling and not consistent 
with the fact that these stars are likely much older (a few hundred million 
years) than their relaxation time (50 to 200~million years).
 
In their most recent analysis Genzel et al.~(2000) also find some evidence that
the fainter `SgrA$^*$ cluster' stars have a tendency for more radial orbits 
(Fig.~7, upper right). This would fit with the fact that the innermost fast 
moving stars (S1, S2 and S8) have orbits with relatively small curvature 
(Genzel et al.~2000, Ghez et al.~2000). Genzel et al.~(2000) propose that the 
SgrA$^*$ cluster stars may be somewhat lower mass (10--20\,M$_\odot$) members 
of the He{\small I}-star cluster which happen to be on plunging, highly 
elliptical orbits and thus make it to the immediate vicinity of SgrA$^*$. 
However, this conclusion must be regarded as tentative and more statistics is 
required for making a conclusive statement.

\section{IS SGRA$^*$ A MASSIVE BLACK HOLE?}
The most exciting aspect of the proper motion data is that they provide 
measurements of stellar velocities in the SgrA$^*$ cluster. Several faint stars
within 0.6$^{\prime\prime}$ of SgrA$^*$ have proper motions in excess of 
1000\,km/s. The fastest star (S1) at a distance of only 
$\sim$0.1$^{\prime\prime}$ ($\sim$800\,AU or 5~light days) from the radio 
source has a proper motion of $\sim$1470\,km/s. The combined radial and proper 
motion data show that stellar velocities increase with a Kepler law 
($\mathrm{v}\sim \mathrm{R}^{-1/2}$) to a scale of $\sim$0.01\,pc. Fig.~8 gives
the present best mass distribution (from Genzel et al.~2000) derived from 
various analyses, including projected mass estimators and Jeans equation 
modeling for both radial and proper motions of the stars. Using the 
(anisotropy-independent) mass estimator proposed by Leonard and Merritt (1989) 
the central mass is $2.7(\pm 0.4)\times 10^6\,$M$_\odot$ for a Galactic center 
distance of $\mathrm{R}_{\mathrm{o}}=8.0\,$kpc (which is also the best estimate 
of $\mathrm{R}_{\mathrm{o}}$ from a comparison of the radial velocity and proper 
motion data, Genzel et al.~2000). Overall the measurements are fitted very well
by a combination of a central point 
mass, plus an extended, near-isothermal stellar cluster with core radius 
$\sim$0.4\,pc and core mass density of 
$4\times 10^6\,\mathrm{M}_\odot/\mathrm{pc}^3$. If the central point mass is 
replaced by a compact dark cluster with a Plummer density distribution, its 
core 
density must exceed 
$3.7\times 10^{12}\,\mathrm{M}_\odot/\mathrm{pc}^3$, almost a million times 
denser than the visible stellar cluster, or the densest globular clusters. The 
Plummer model also requires that such a dark cluster would have to have a very 
steep density distribution outside of its core radius 
($\mathrm{density}\sim \mathrm{R}^{-4\ldots -5}$, for 
$\mathrm{R}>\mathrm{R}_{\mathrm{core}}\sim 5.8$\,milli-parsec), very different 
from an isothermal distribution. The mass to bolometric luminosity ratio of 
this central dark mass thus is a few hundred (in solar units) or greater.

Simple physical considerations show that clusters of low mass stars 
(e.g.~white dwarfs), neutron stars, stellar black holes or sub-stellar entities
(e.g.~brown dwarfs, rocks) with the observed properties of the dark mass cannot
be stable for longer than $\sim10^7\,$years (Maoz 1995, 1998, Genzel et 
al.~1997). It is also not possible that the dark mass concentration is the very
dense (=\,core-collapsed) state of a dynamically evolving cluster of above 
objects. In that 
case the distribution --- while very dense in its very small core --- would 
have a soft, quasi-isothermal envelope, unlike what is observed in the Galactic
Center (Genzel et al.~1997). The most likely configuration of the dark Galactic
Center mass distribution thus is a black hole. Maoz (1998) points out that the 
only --- albeit highly improbable --- alternatives to a massive black hole are 
a concentration of heavy bosons and a compact cluster of light 
($<$0.005\,M$_\odot$) black holes. 

Two further arguments substantially strengthen the conclusion that the dark 
mass in the Galactic Center in fact must be a massive black hole. The first 
comes from the fact that SgrA$^*$ itself is known from VLBI measurements to 
have a proper motion less than about 20\,km/s relative to the Galactic Center 
reference frame (Backer and Sramek 1999, Reid et al.~1999). Hence the two order
of magnitude difference in velocities between the radio source and the nearby 
SgrA$^*$ cluster stars means that SgrA$^*$ must have a mass 
$\gg$$10^3\,$M$_\odot$ (Reid et al.~1999), unless its true motion is exactly 
along the line of sight (Genzel et al.~1997). If one further assumes that the 
mass of SgrA$^*$ must be at least as concentrated as its radio emission 
(1\,AU corresponds to 17~Schwarzschild radii of a 3~million solar mass black 
hole), the inferred density of SgrA$^*$ must be 
$>$$10^{18}\,\mathrm{M}_\odot/\mathrm{pc}^3$. The second argument is an 
inversion of the well known dilemma that if SgrA$^*$ is a three million solar 
mass black hole it is currently radiating at a rest mass energy to radiation, 
conversion efficiency of only $10^{-5}$ to $10^{-6}$, considering the accretion
of stellar wind gas from its environment (Melia 1992, Genzel et al.~1994). The 
only possible way for explaining this feeble emission (other than very large 
time variability of the accretion) is the argument that in purely radial 
(Bondi-Hoyle) or in low density, non-radial flows most of the rest mass energy 
of the accretion flow can be advected into the hole, rather than radiated away 
(Rees et al.~1982, Melia 1992, 1994, Narayan et al.~1995, 1998). This 
explanation, however, requires the existence of an event horizon and does not 
work with any configuration but a black hole (Narayan et al.~1998). Taking all 
these arguments together it is hard to escape the conclusion that the core of 
the Milky Way in fact harbors a million solar mass, central black hole.

Further progress can be expected soon. The first detections of curvature in the
trajectories of S1, S2 and S8 from the Keck proper motion experiment (Ghez et 
al.~2000) demonstrate that SgrA$^*$ indeed is near the focus of the orbits and 
that fairly accurate orbits for individual stars can be determined over the next
few years. Further details of the mass distribution (single massive black hole?,
halo of massive objects surrounding the central mass?) will then come from the 
precision analysis of individual orbits, rather than from the present coarse, 
statistical tools. Deep diffraction limited, adaptive optics imaging and 
spectroscopy will result in observations of more stars and of even faster 
moving stars within $<$0.1$^{\prime\prime}$ of SgrA$^*$ (cusp?). The AO imaging
also will probe deeper down the main sequence and make feasible the detection 
of gravitational microlensing of background stars by the central black hole 
(Alexander and Sternberg 1999). Imaging spectroscopy will lead to a better 
understanding of the evolution of this unique stellar cluster.

\vspace{\baselineskip}
\noindent{\Large\bf REFERENCES}
\vspace{\baselineskip}
\par\noindent\hangindent=\parindent\hangafter1\em\rm
Alexander, T. 1999, Ap.J. 527, 835
\par\noindent\hangindent=\parindent\hangafter1\em\rm
Alexander, T. and Sternberg, A. 1999, Ap.J. 520, 137
\par\noindent\hangindent=\parindent\hangafter1\em\rm
Allen, D.A. and Sanders, R.H. 1986, NATURE 319, 191
\par\noindent\hangindent=\parindent\hangafter1\em\rm
Allen, D.A., Hyland, A.R. and Hillier, D.J. 1990, MNRAS 244, 706
\par\noindent\hangindent=\parindent\hangafter1\em\rm
Allen, D.A. 1994, in "The Nuclei of Normal Galaxies", eds. R.Genzel and A.Harris 
   (Dordrecht:Kluwer),  293 
\par\noindent\hangindent=\parindent\hangafter1\em\rm
Backer, D.C. and Sramek, R.A. 1999, Ap.J. 524, 805
\par\noindent\hangindent=\parindent\hangafter1\em\rm
Baganoff et al. 2000, in prep.
\par\noindent\hangindent=\parindent\hangafter1\em\rm
Blum, R.D., Sellgren, K. and DePoy, D.L. 1996a , A.J. 112, 1988
\par\noindent\hangindent=\parindent\hangafter1\em\rm
Blum, R.D., Sellgren, K. and DePoy, D.L. 1996b , Ap.J. 470, 864
\par\noindent\hangindent=\parindent\hangafter1\em\rm
Blum, R.D., dePoy, D.L. and Sellgren, K.1995b, Ap.J.441, 603 
\par\noindent\hangindent=\parindent\hangafter1\em\rm
Blum, R.D., Sellgren, K. and dePoy, D.L..1995a ,Ap.J.440, L17 
\par\noindent\hangindent=\parindent\hangafter1\em\rm
Bower, G.C. and Backer, D.C> 1998, Ap.J. 496, L97
\par\noindent\hangindent=\parindent\hangafter1\em\rm
Carr, J.S., Sellgren, K. and Balachandian, S.C. 1999, submitted (astro-ph 9909037)
\par\noindent\hangindent=\parindent\hangafter1\em\rm
Cotera, A.S., Erickson, E.F., Colgan, S.W.J., Simpson, J.P., Allen D.A. and Burton, M.G. 1996, Ap.J. 461, 750 
\par\noindent\hangindent=\parindent\hangafter1\em\rm
Davidge, T.J., Simons, D.A>, Rigaut, F., Doyon, R. and Crampton, D. 1997, A.J. 114, 2586
\par\noindent\hangindent=\parindent\hangafter1\em\rm
Eckart, A., Ott, T. and Genzel, R. 1999, Astr.Ap. 352, L22
\par\noindent\hangindent=\parindent\hangafter1\em\rm
Eckart, A. and Genzel, R. 1997, MNRAS 284, 576
\par\noindent\hangindent=\parindent\hangafter1\em\rm
Eckart, A. and Genzel, R.1996, NATURE 383, 415
\par\noindent\hangindent=\parindent\hangafter1\em\rm
Eckart, A. Genzel, R., Hofmann, R., Sams, B.J. and Tacconi-Garman, L.E. 1995, Ap.J. 445, L26 
\par\noindent\hangindent=\parindent\hangafter1\em\rm
Eckart, A. Genzel, R., Hofmann, R., Sams, B.J. and Tacconi-Garman, L.E. 1993, Ap.J. 407, L77
\par\noindent\hangindent=\parindent\hangafter1\em\rm
Figer, D. et al. 2000, Ap.J. in press (astro-ph 0001171)
\par\noindent\hangindent=\parindent\hangafter1\em\rm
Figer, D., McLean, I. And Morris, M. 1999a, Ap.J. 514, 202
\par\noindent\hangindent=\parindent\hangafter1\em\rm
Figer, D., Kim, S.S., Morris, M., Serabyn, E., Rich, R.M., and McLean, I. 1999b, Ap.J. 525, 750
\par\noindent\hangindent=\parindent\hangafter1\em\rm
Figer, D. F., McLean, I.S. and Morris, M. 1995, Ap.J. 447, L29 
\par\noindent\hangindent=\parindent\hangafter1\em\rm
Forrest, W.J., Shure, M.A., Pipher, J.L. and Woodward, C.A 1987, .in "The Galactic Center", ed.D.Backer, AIP Conf. Proc 155, 153 
\par\noindent\hangindent=\parindent\hangafter1\em\rm
Genzel, R., Pichon, C., Eckart, A., Gerhard, O., Ott, T. 2000, MNRAS in press (astro-ph 0001428)
\par\noindent\hangindent=\parindent\hangafter1\em\rm
Genzel, R. and Eckart, A. 1998, C.R.Acad.Sci.Ser.II 326, 69
\par\noindent\hangindent=\parindent\hangafter1\em\rm
Genzel, R., Eckart, A., Ott, T. and Eisenhauer, F. 1997, MNRAS 291, 219
\par\noindent\hangindent=\parindent\hangafter1\em\rm
Genzel, R., Thatte, N., Krabbe, A., Kroker, H. and Tacconi-Garman, L.E. 1996, Ap.J.472, 153
\par\noindent\hangindent=\parindent\hangafter1\em\rm
Genzel, R., Hollenbach, D.J. and Townes, C.H. 1994, Rep.Progr.Phys. 57, 417
\par\noindent\hangindent=\parindent\hangafter1\em\rm
Genzel, R. and Townes, C.H. 1987, Ann.Rev.Astr.Ap. 25, 377
\par\noindent\hangindent=\parindent\hangafter1\em\rm
Gerhard, O. 2000, in prep.
\par\noindent\hangindent=\parindent\hangafter1\em\rm
Ghez, A. et al. 2000, in prep.
\par\noindent\hangindent=\parindent\hangafter1\em\rm
Ghez, A., Becklin, E., Morris, M. and Klein, B. 1998, Ap.J. 509, 678
\par\noindent\hangindent=\parindent\hangafter1\em\rm
Haller, J.W. and Rieke, M.J. 1989,in "The Center of the Galaxy" , ed. M.Morris, (Dordrecht:Kluwer), 487 
\par\noindent\hangindent=\parindent\hangafter1\em\rm
Haller, J.W., Rieke, M.J., Rieke, G.H., Tamblyn, P., Close, L. and Melia, F. 1986, Ap.J. 456, 194
\par\noindent\hangindent=\parindent\hangafter1\em\rm
Hanson, M.M., Conti, P.S. and Rieke, M.J. 1996, Ap.J.Suppl.107, 281
\par\noindent\hangindent=\parindent\hangafter1\em\rm
Heger, A. and Langer, N. 2000, submitted Astr.Ap. (astro-ph 0005110)
\par\noindent\hangindent=\parindent\hangafter1\em\rm
Koyama, K., Maeda, Y., Sonobe, T., Takeshima, T., Tanaka, Y. and Yamauchi, S. 1996, PASJ 48, 249
\par\noindent\hangindent=\parindent\hangafter1\em\rm
Krabbe, A. et al. 1995, Ap.J. 447, L95
\par\noindent\hangindent=\parindent\hangafter1\em\rm
Krabbe, A., Genzel, R., Drapatz, S. and Rotatciuc, V. 1991, Ap.J. 382, L19
\par\noindent\hangindent=\parindent\hangafter1\em\rm
Krichbaum, T.P., Witzel, A. and Zensus, J.A. in 'The Central Parsecs of the Galaxy', eds. H.Falcke, A.Cotera, W.Duschl, F.Melia and M.Rieke, ASP Conf.Series Vol. 186 (ASP: San Francisco), 89
\par\noindent\hangindent=\parindent\hangafter1\em\rm
Lacy, J.H., Townes, C.H. and Hollenbach, D.J.1982 J. 262, 120 
\par\noindent\hangindent=\parindent\hangafter1\em\rm
Lacy, J.H., Townes, C.H., Geballe, T.R. and Hollenbach, D.J. 1980, Ap.J. 241, 132
\par\noindent\hangindent=\parindent\hangafter1\em\rm
Langer, N. and Maeder, A. 1995, Astr.Ap. 295, 685
\par\noindent\hangindent=\parindent\hangafter1\em\rm
Lee, H.M. 1994, in "The Nuclei of Normal Galaxies", eds. R.Genzel and A.I. Harris (Dordrecht: Kluwer), 335
\par\noindent\hangindent=\parindent\hangafter1\em\rm
Leonard, P.J.T. and Merritt, D. 1989, Ap.J. 339, 195
\par\noindent\hangindent=\parindent\hangafter1\em\rm
Libonate, S., Pipher, J.L., Forrest, W.J. and Ashby, M.L.N.1995 , Ap.J. 439, 202 
\par\noindent\hangindent=\parindent\hangafter1\em\rm
Lindqvist, M., Habing, H. and Winnberg, A. 1992, Astr.Ap. 259, 118
\par\noindent\hangindent=\parindent\hangafter1\em\rm
Lo, K.Y., Shen, Z.-Q., Zhao, J.H., Ho, P.T.P. in 'The Central Parsecs of the Galaxy', eds. H.Falcke, A.Cotera, W.Duschl, F.Melia and M.Rieke, ASP Conf.Series Vol. 186 (ASP: San Francisco), 72
\par\noindent\hangindent=\parindent\hangafter1\em\rm
Lutz, D. 1999, in The Universe as seen by ISO, eds.P.Cox and M.F.Kessler, ESASP 427 (ESA: Nordwijk), 623
\par\noindent\hangindent=\parindent\hangafter1\em\rm
Lutz, D. et al. 1996, Astr.Ap. 315, L269
\par\noindent\hangindent=\parindent\hangafter1\em\rm
Lynden-Bell, D. and Rees, M. 1971 , MNRAS 152, 461 
\par\noindent\hangindent=\parindent\hangafter1\em\rm
Maoz, E. 1998, Ap.J. 494, L131
\par\noindent\hangindent=\parindent\hangafter1\em\rm
Maoz, E. 1995, Ap.J. 447, L91
\par\noindent\hangindent=\parindent\hangafter1\em\rm
McGinn, M.T., Sellgren, K., Becklin, E.E. and Hall, D.N.B. 1989, Ap.J. 338, 824
\par\noindent\hangindent=\parindent\hangafter1\em\rm
Melia, F. 1992, Ap.J. 387, L25
\par\noindent\hangindent=\parindent\hangafter1\em\rm
Melia, F. 1994, Ap.J. 426, 577
\par\noindent\hangindent=\parindent\hangafter1\em\rm
Menten, K.M., Eckart, A., Reid, M.J. and Genzel, R. 1997, Ap.J. 475, L111
\par\noindent\hangindent=\parindent\hangafter1\em\rm
Meynet, G. et al. 1994, Astr.Ap.(Suppl.) 103, 97
\par\noindent\hangindent=\parindent\hangafter1\em\rm
Mezger, P.G., Duschl, W.J. and Zylka, R. 1996, Astr.Ap. Rev. 7, 289
\par\noindent\hangindent=\parindent\hangafter1\em\rm
Morris, M., Maillard, J.-P. et al. 2000, in prep.
\par\noindent\hangindent=\parindent\hangafter1\em\rm
Morris, M. and Serabyn, E.1996, Ann.Rev.Astr.Ap. 34, 645
\par\noindent\hangindent=\parindent\hangafter1\em\rm
Morris, M. 1993, Ap.J. 408, 496
\par\noindent\hangindent=\parindent\hangafter1\em\rm
Morris, P.W., Eenens, P.R.J., Hanson, M.M., Conti, P.S., Blum, R.D. 1996, Ap.J. 470, 597
\par\noindent\hangindent=\parindent\hangafter1\em\rm
Najarro, F., Hillier, D.J., Figer, D. and Geballe, T.R. 1999, in 'The Central Parsecs of the Galaxy', eds. H.Falcke, A.Cotera, W.Duschl, F.Melia and M.Rieke, ASP Conf.Series Vol. 186 (ASP: San Francisco), 340
\par\noindent\hangindent=\parindent\hangafter1\em\rm
Najarro, F., Krabbe, A., Genzel, R., Lutz, D., Kudritzki, R.P.and Hillier, D.J. 1997, Astr.Ap.325, 700 
\par\noindent\hangindent=\parindent\hangafter1\em\rm
Najarro, F. et al.1994 ,  Astr.Ap. 285, 573 
\par\noindent\hangindent=\parindent\hangafter1\em\rm
Narayan, R., Mahadevan, R., Grindlay, J., Popham, R.G. and Gammie, C. 1998, Ap.J. 492, 554
\par\noindent\hangindent=\parindent\hangafter1\em\rm
Narayan, R., Yi, I. and Mahadevan, R. 1995, NATURE 374, 623
\par\noindent\hangindent=\parindent\hangafter1\em\rm
Ott, T. et al. 2000, in prep.
\par\noindent\hangindent=\parindent\hangafter1\em\rm
Ott, T., Eckart, A. and Genzel 1999, Ap.J. 523, 248
\par\noindent\hangindent=\parindent\hangafter1\em\rm
Phinney, E.S. 1989, in "The Center of the Galaxy", ed.M.Morris (Kluwer:Dordrecht), 543
\par\noindent\hangindent=\parindent\hangafter1\em\rm
Ramirez, S.V., Sellgren, K., Carr, J.S., Balachandran, S.C., Blum, R., Terndrup, D.M. and Steed, A. 2000, submitted (astro-ph 0002062)
\par\noindent\hangindent=\parindent\hangafter1\em\rm
Rees, M., Phinney, E.S., Begelman, M.C. and Blandford, R.D. 1982, NATURE 295, 17
\par\noindent\hangindent=\parindent\hangafter1\em\rm
Reid, M.J., Readhead, A.C.S., Vermeulen, R.C. and Treuhaft, R.N. 1999, Ap.J. 524, 816
\par\noindent\hangindent=\parindent\hangafter1\em\rm
Rieke, G.H. and Rieke, M.J.1994 in "The Nuclei of Normal Galaxies", eds. R.Genzel and A. Harris, 283 
\par\noindent\hangindent=\parindent\hangafter1\em\rm
Rieke, G.H. and Lebofsky, M.J. 1982, in "The Galactic Center", eds. G.Riegler and R.D.Blandford, AIP conf. proc. 83 (New York), 194 
\par\noindent\hangindent=\parindent\hangafter1\em\rm
Rieke, G.H. and Rieke, M.J. 1988, Ap.J. 330, L33
\par\noindent\hangindent=\parindent\hangafter1\em\rm
Rieke, M.J.1999, in 'The Central Parsecs of the Galaxy', eds. H.Falcke, A.Cotera, W.Duschl, F.Melia and M.Rieke, ASP Conf.Series Vol. 186 (ASP: San Francisco), 32
\par\noindent\hangindent=\parindent\hangafter1\em\rm
Sellgren, K., McGinn, M.T., Becklin, E.E. and Hall, D.N.B. 1990, Ap.J. 359, 112
\par\noindent\hangindent=\parindent\hangafter1\em\rm
Shields, J.C. and Ferland, G.J. 1994, Ap.J. 430, 236
\par\noindent\hangindent=\parindent\hangafter1\em\rm
Sjouwerman, L.O., Habing, H.J., Lindqvist, M., H.J. van Langenvelde and A.Winnberg in 'The Central Parsecs of the Galaxy', eds. H.Falcke, A.Cotera, W.Duschl, F.Melia and M.Rieke, ASP Conf.Series Vol. 186 (ASP: San Francisco), 379
\par\noindent\hangindent=\parindent\hangafter1\em\rm
Stolovy, S.R., McCarthy, D.W., Melia, F., Rieke, G.H., Rieke, M.J. and Yusef-Zadeh, F. 1999, in 'The Central Parsecs of the Galaxy', eds. H.Falcke, A.Cotera, W.Duschl, F.Melia and M.Rieke, ASP Conf.Series Vol. 186 (ASP: San Francisco), 39
\par\noindent\hangindent=\parindent\hangafter1\em\rm
Sunyaev, R., Markevitch, M. and Pavlinsky, M. 1993, Ap.J. 407, 606
\par\noindent\hangindent=\parindent\hangafter1\em\rm
Stolovy, S. R., Hayward, T.L. and Herter, T. 1996, Ap.J. 470, L45
\par\noindent\hangindent=\parindent\hangafter1\em\rm
Tamblyn, P., Rieke, G.H., Hanson, M.M., Close, L.M., McCarthy, D. W. and Rieke, M.J. 1996, Ap.J. 456, 206
\par\noindent\hangindent=\parindent\hangafter1\em\rm
Tanner, A.M., Ghez, A., Morris, M. and Becklin, E. 1999, in 'The Central Parsecs of the Galaxy', eds. H.Falcke, A.Cotera, W.Duschl, F.Melia and M.Rieke, ASP Conf.Series Vol. 186 (ASP: San Francisco), 351
\par\noindent\hangindent=\parindent\hangafter1\em\rm
Telesco, C.M., Davidson, J.A. and Werner, M.W. 1996, Ap.J. 456, 541
\par\noindent\hangindent=\parindent\hangafter1\em\rm
Wollman, E.R., Geballe, T.R., Lacy, J.H., Townes, C.H. and Rank, D.M. 1977, Ap.J. 218, L103
\newpage
\par\noindent{\Large\bf FIGURE CAPTIONS}
\vspace{\baselineskip}
\par\noindent
{\bf Fig.~1} Grey scale K-band image of the central 
$\sim$19$^{\prime\prime}$ (0.8\,pc) of the Galactic Center, obtained with 
speckle imaging at 0.15$^{\prime\prime}$ resolution at the 3.5m ESO NTT 
(Eckart et al.~1993, 1995, Eckart and Genzel 1997). The right inset shows the 
central SgrA$^*$ cluster of faint (`SgrA$^*$ cluster') stars in the immediate 
vicinity of the compact radio source SgrA$^*$ (cross, positioning and error 
bars from Menten et al.~(1997), as obtained with 0.05$^{\prime\prime}$ speckle 
imaging on the 10m Keck telescope (Ghez et al.~1998). 

\vspace{\baselineskip}
\par\noindent
{\bf Fig.~2} K-band spectrum ($\mathrm{R}=2000$) of IRS13E (left) and of an 
average of IRS16NE, C, NW and SW (right) obtained with the MPE 3D spectrometer 
on the ESO-MPG 2.2m telescope on LaSilla (Genzel et al.~2000, Ott et al.~2000).
Wavelengths of important transitions/species are marked by arrows. IRS13E has a
spectrum characterstic of a late WN or WC star. The IRS16 stars have spectra 
similar to Luminous Blue Variables (LBVs, like AG~Car, P-Cyg or $\eta$~Car),or 
ON~stars (Tamblyn et al.~1996).

\vspace{\baselineskip}
\par\noindent
{\bf Fig.~3} Hertzsprung-Russell diagram of stars in the central parsec. With 
the exception of the location of the `SgrA$^*$ cluster' (large circle with long
arrow), all stars entering this diagram have 
$\mathrm{m}(\mathrm{K})<13$. Each star is marked as a filled circle. 
Temperatures and luminosities of the late type stars are 
derived from K-band spectroscopy and K-magnitudes (Ott et al.~2000, Blum et 
al.~1996a). Temperatures and luminosities of the early type stars 
(He{\small I}-stars) are derived from the non-LTE modelling of Najarro et 
al.~(1994, 1997). Identifications of a few He{\small I}-stars and the red 
supergiant IRS7 are given. Several very cold objects have featureless K-band 
spectra (see 
Krabbe et al.~1995) with strong long-wavelength dust excess, indicating that 
they may be dust enshrouded young stellar objects, or dusty massive stars. 
Heavy and dashed heavy lines denote the main sequence and giant branches for 
solar and twice solar metallicity, respectively, with masses marked. Stellar 
tracks for twice solar metallicity from the work of Meynet et al.~(1994) are 
plotted for 4~different masses.

\vspace{\baselineskip}
\par\noindent
{\bf Fig.~4} VLT-ISAAC spectroscopy (0.6$^{\prime\prime}$ slit) of the 
SgrA$^*$ cluster stars (Eckart et al.~1999, see also Genzel et al.~1997, Figer 
et al.~2000). The SgrA$^*$ cluster stars have featureless K-band spectra, 
indicating that their temperature is greater than about 5000\,K (large circle 
and arrow in Fig.~3). If they are main sequence stars they are of type 
O9--B2 and of mass 10 to 20~solar masses.

\vspace{\baselineskip}
\par\noindent
{\bf Fig.~5} High resolution ($\mathrm{R}\sim 40,000$, CSHELL on IRTF) H-band 
spectra of three supergiant stars of similar spectra type, IRS7 in the Galactic
Center (upper histogram), VV~Cep (lower histogram) and $\alpha$~Ori 
(continuous line overplotted on both histograms) (from Carr et al.~2000, see 
also Ramirez et al.~2000). VV~Cep and $\alpha$~Ori have near solar abundances.

\vspace{\baselineskip}
\par\noindent
{\bf Fig.~6} Proper motion vectors of early type stars (lighter grey arrows: 
He{\small I} emission line stars and brighter members of IRS16 cluster), late 
type stars (darker grey arrows) and of SgrA$^*$ cluster stars (right inset). 
The data are from Genzel et al.~2000 which include the most recent NTT proper 
motion results, as well as the Keck results from Ghez et al.~1998. The fastest 
star (S1 near SgrA$^*$) has a velocity of $1470\pm 100$\,km/s.

\vspace{\baselineskip}
\par\noindent
{\bf Fig.~7} The anisotropy measure 
$\gamma_{\mathrm{RT}}=(\mathrm{v}_{\mathrm{T}}^2-\mathrm{v}_{\mathrm{R}}^2)/
(\mathrm{v}_{\mathrm{T}}^2+\mathrm{v}_{\mathrm{R}}^2$)for different sub-samples
of proper motion stars. Here $\mathrm{v}_{\mathrm{T}}$ and 
$\mathrm{v}_{\mathrm{R}}$ are the sky-projected tangential and radial 
components of the proper motion of a given star. Bottom right: all stars with 
proper motions at $\mathrm{R}<5^{\prime\prime}$ from SgrA$^*$. Bottom left: 
Late type stars only. Upper left inset: He{\small I} emission line stars only. 
Upper right inset: Stars in the SgrA$^*$ cluster 
($\mathrm{R}<0.8^{\prime\prime}$).

\vspace{\baselineskip}
\par\noindent
{\bf Fig.~8} Mass distribution in the central 10\,pc of the Galaxy, as obtained
from stellar (and gas) dynamics. Shown as filled rectangles (with typical 
1$\sigma$ error bars for two points) are mass estimates from Jeans equation 
analysis and projected mass estimators, obtained from proper motions and radial
motions and  assuming a Sun-Galactic center distance of 8\,kpc (Genzel et 
al.~1997, 2000, Ghez et al.~1998). The grey curve with error bars is a Jeans 
analysis including anisotropy (Genzel et al.~2000). The thick dashed curve 
represents the mass model for the (visible) stellar cluster 
($\mathrm{M}/\mathrm{L}_{2~\mathrm{micron}}=2$, 
$\mathrm{R}_{\mathrm{core}}=0.38\,$pc, 
$\rho_{\mathrm{R}=0}=4\times 10^6\,\mathrm{M}_\odot/pc^3$, Genzel et al.~1996).
The thick continuous curve is the sum of this stellar cluster, plus a point 
mass of $2.9\times 10^6\,\mathrm{M}_\odot$. The thin dotted curve is the sum of
 the 
visible stellar cluster, plus an $a=5$ Plummer model of a 
dark cluster of central density 
$4\times 10^{12}\,\mathrm{M}_\odot/\mathrm{pc}^3$ and 
$\mathrm{R}_{\mathrm{o}}=0.0065\,$pc.
\newpage
\centerline{\psfig{file=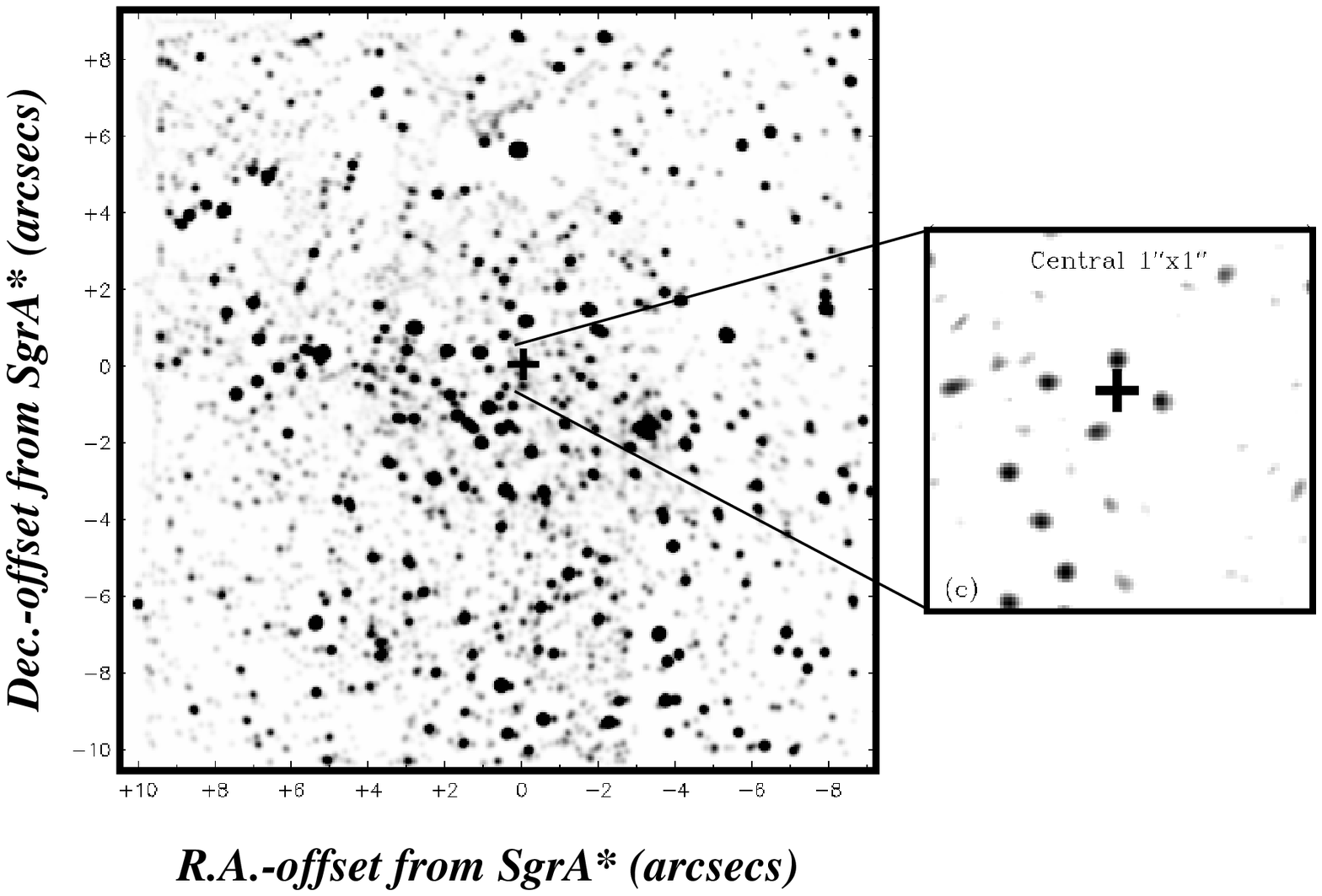}}
\newpage
\centerline{\psfig{file=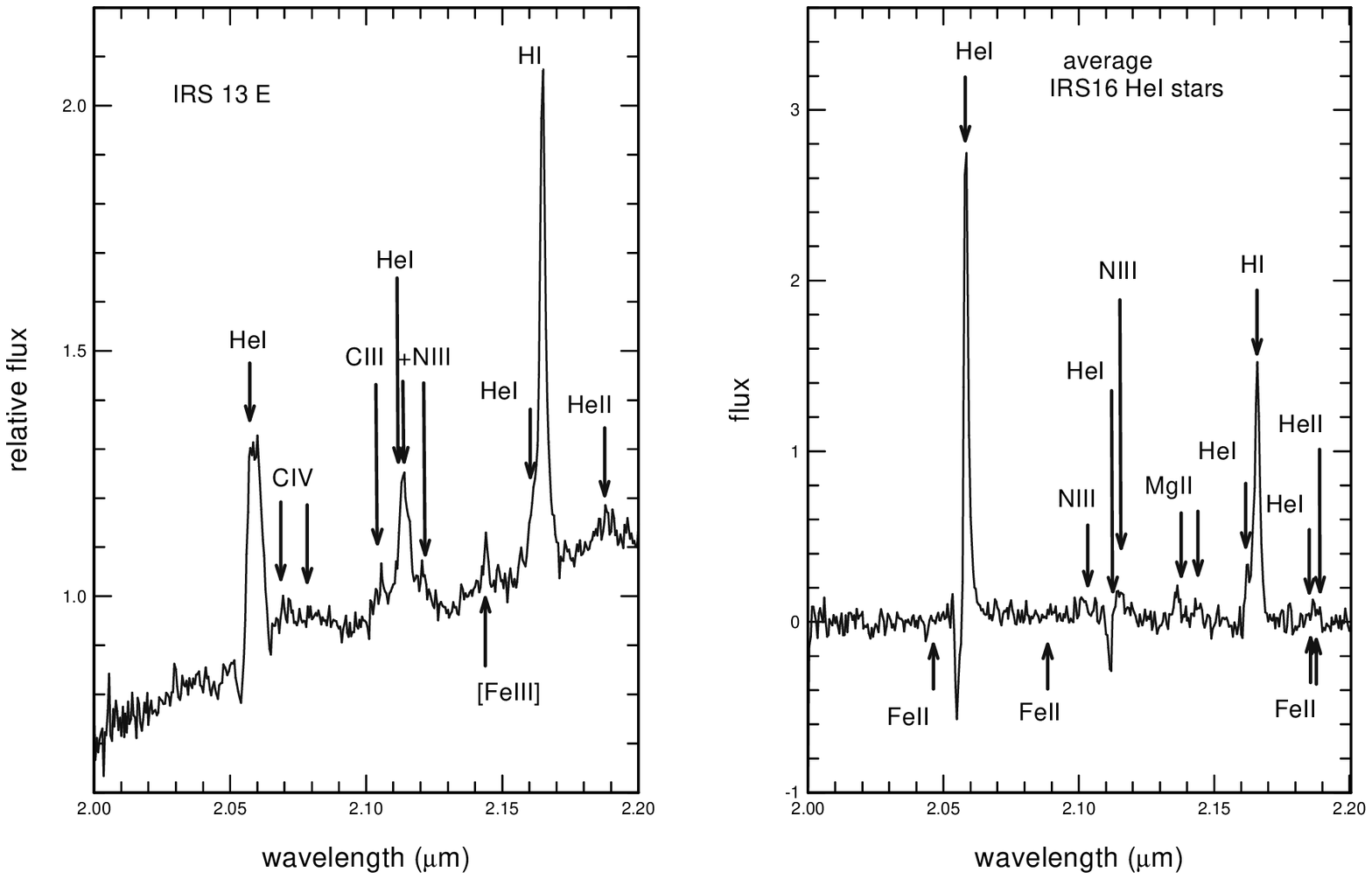}}
\newpage
\centerline{\psfig{file=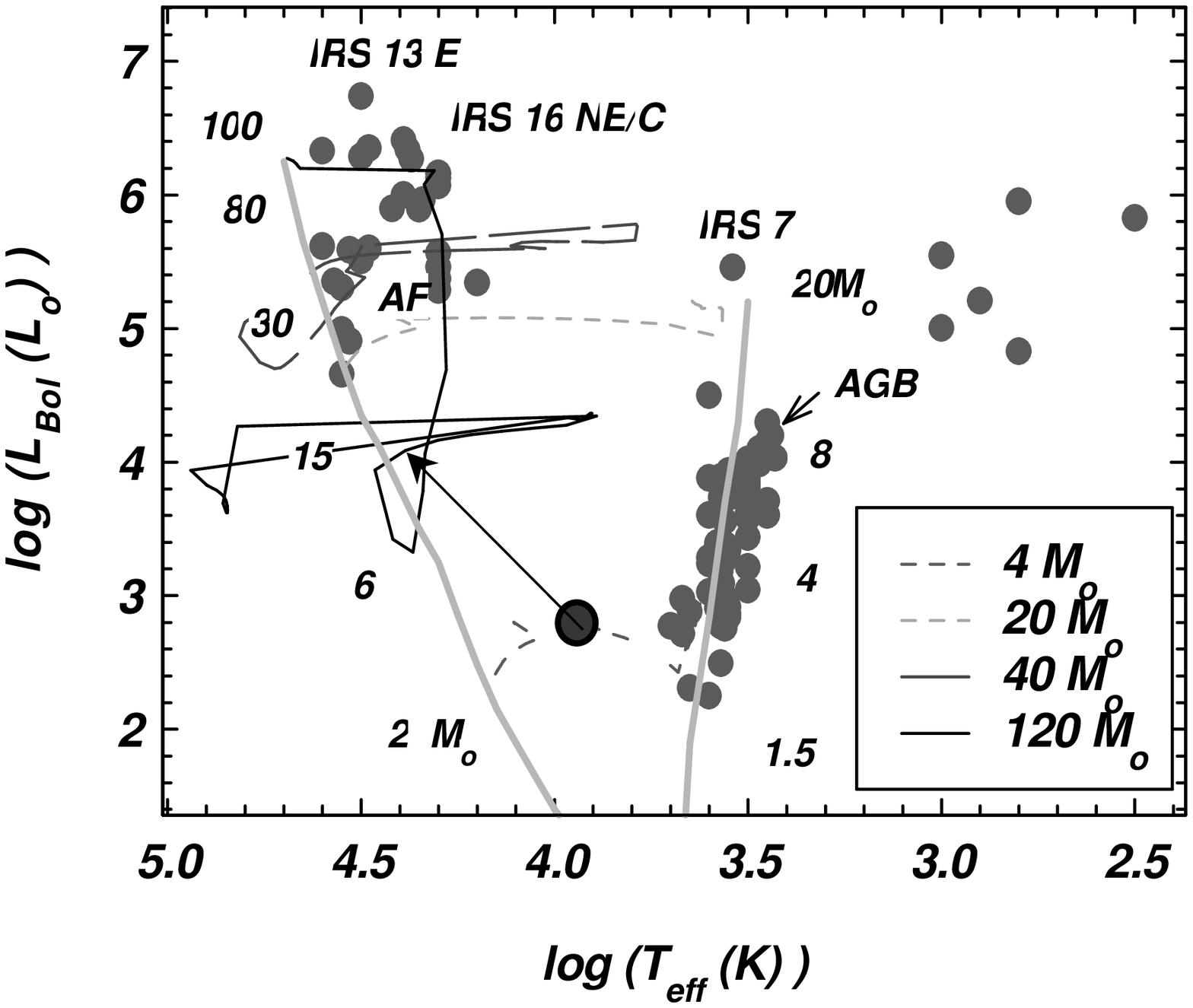}}
\newpage
\centerline{\psfig{file=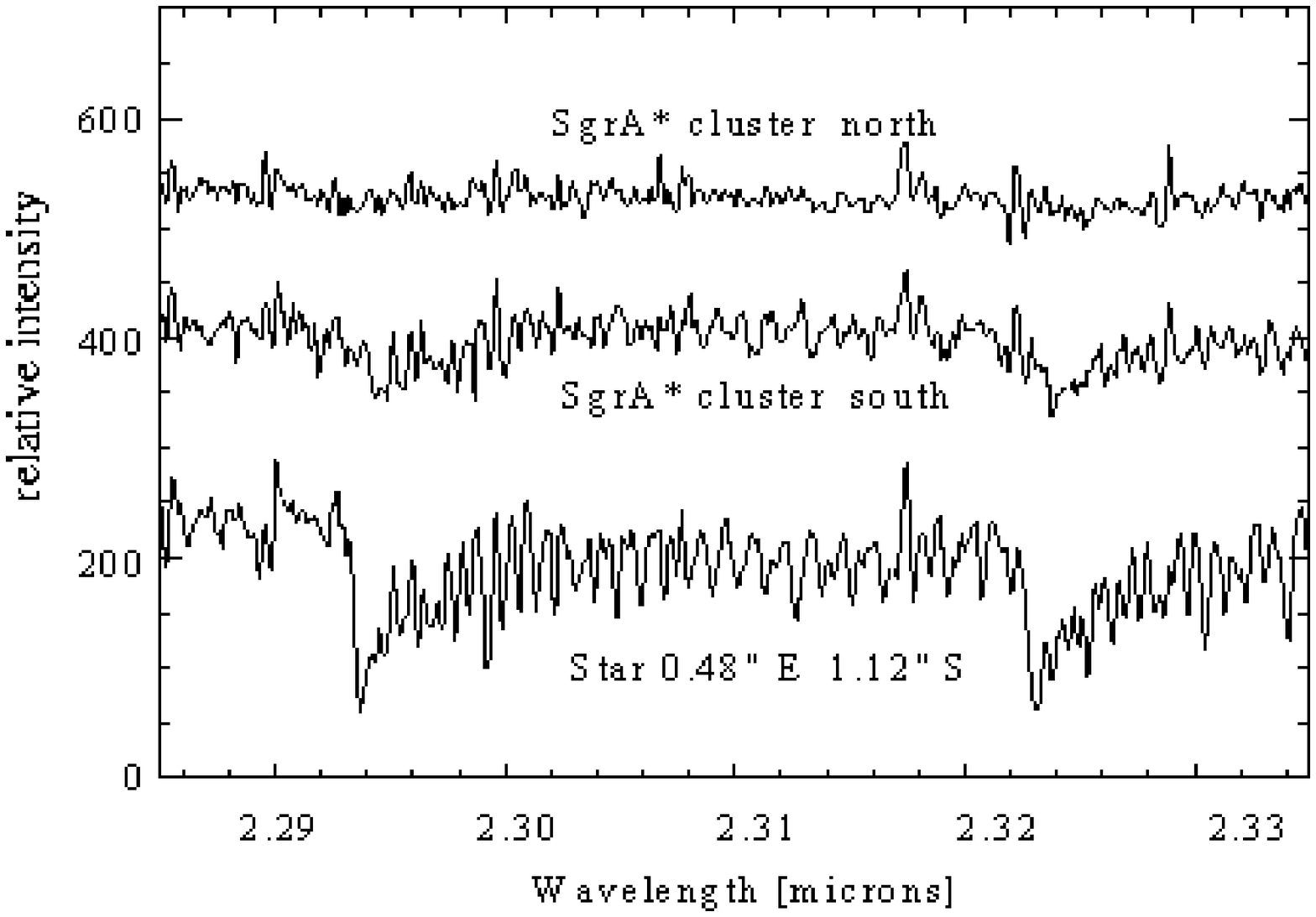}}
\newpage
\centerline{\psfig{file=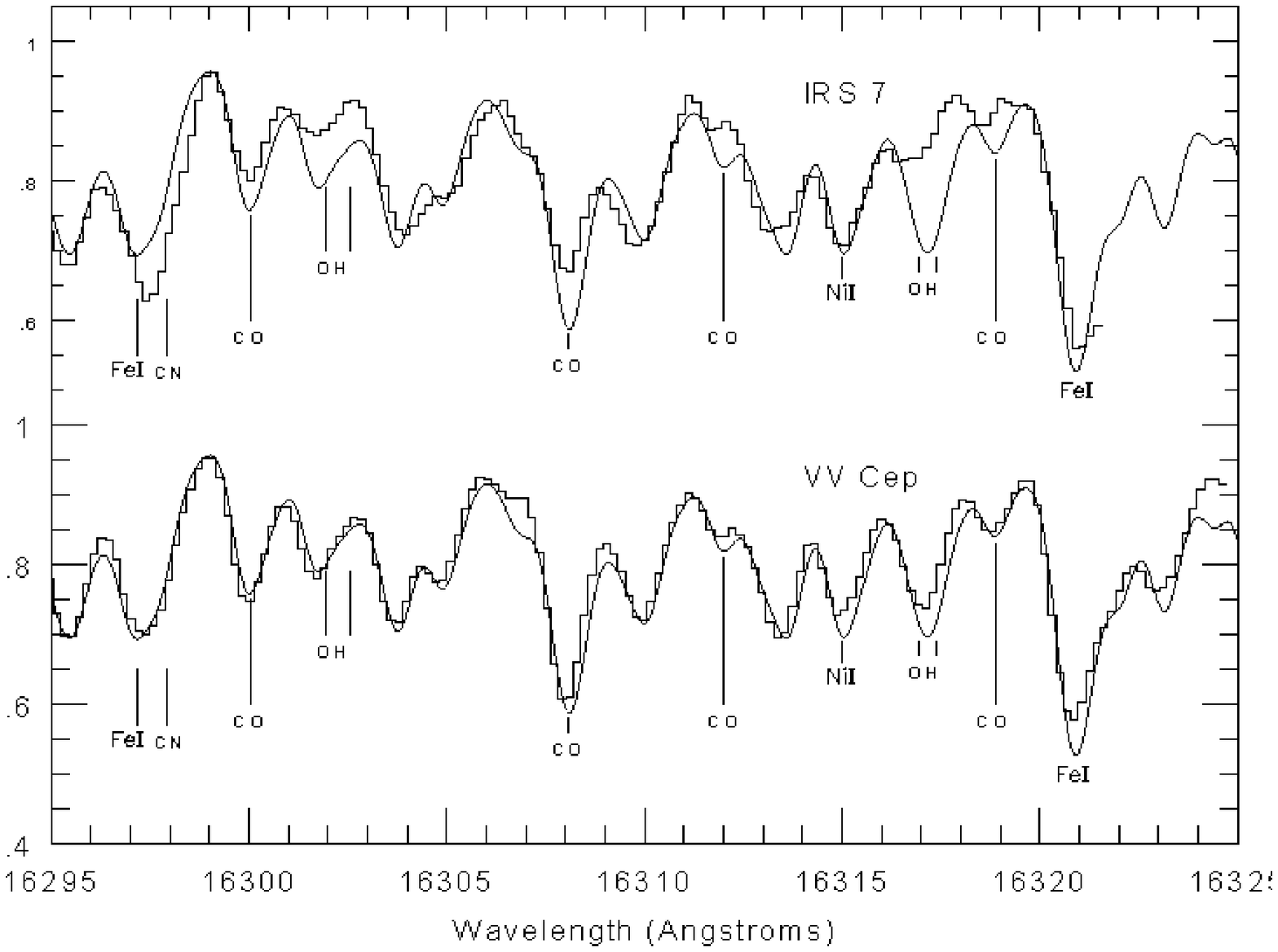}}
\newpage
\centerline{\psfig{file=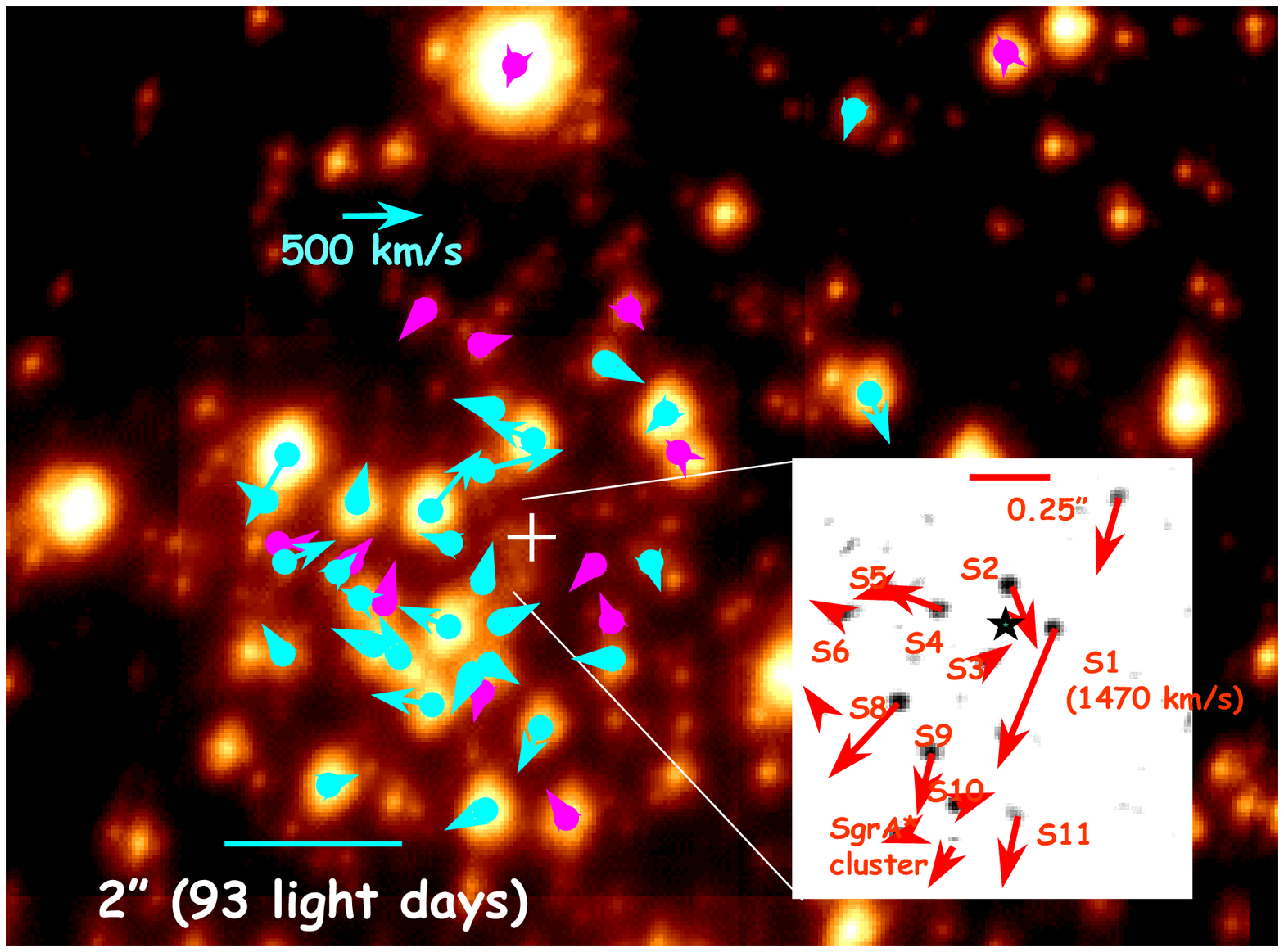}}
\newpage
\centerline{\psfig{file=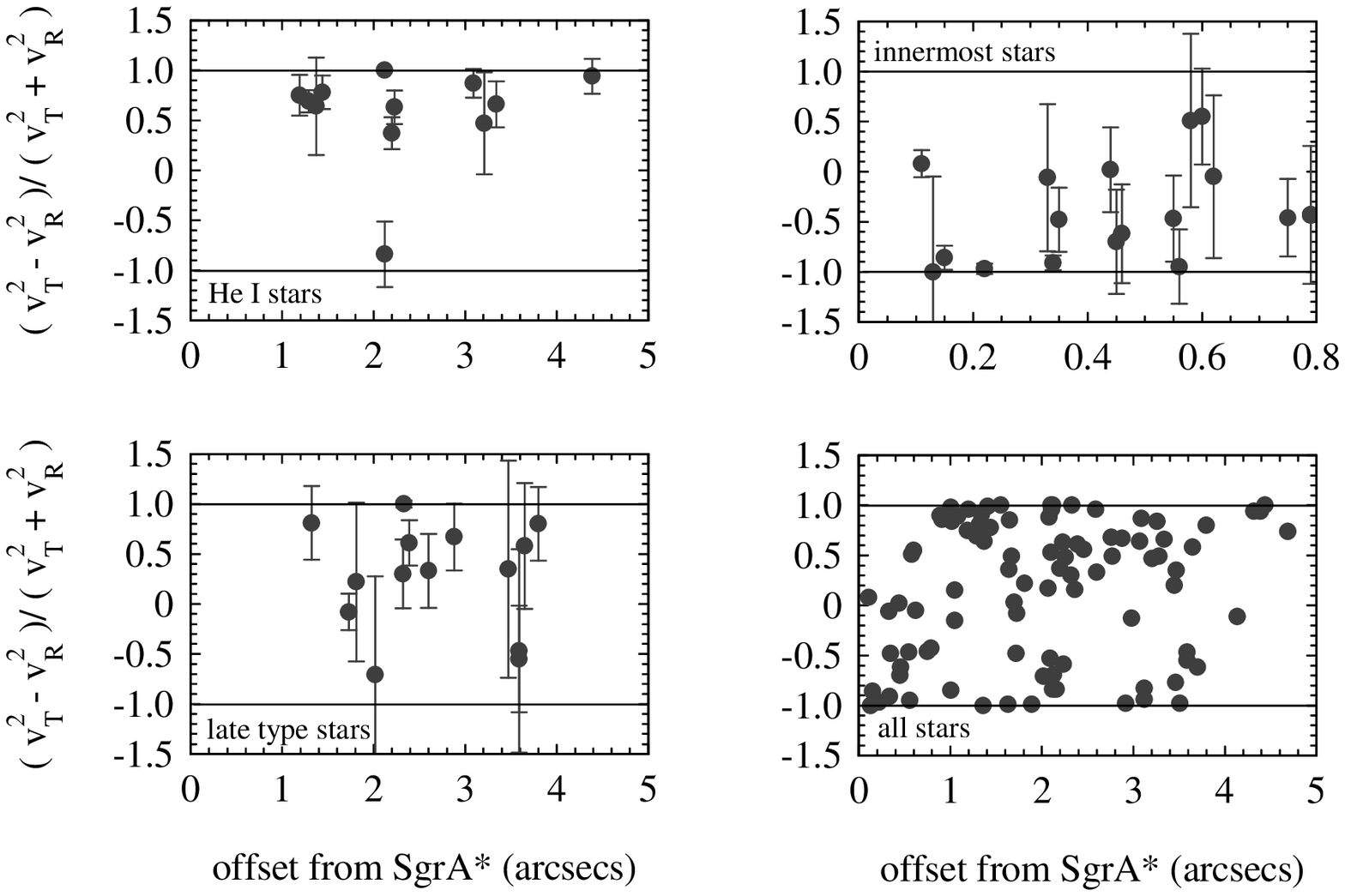}}
\newpage
\centerline{\psfig{file=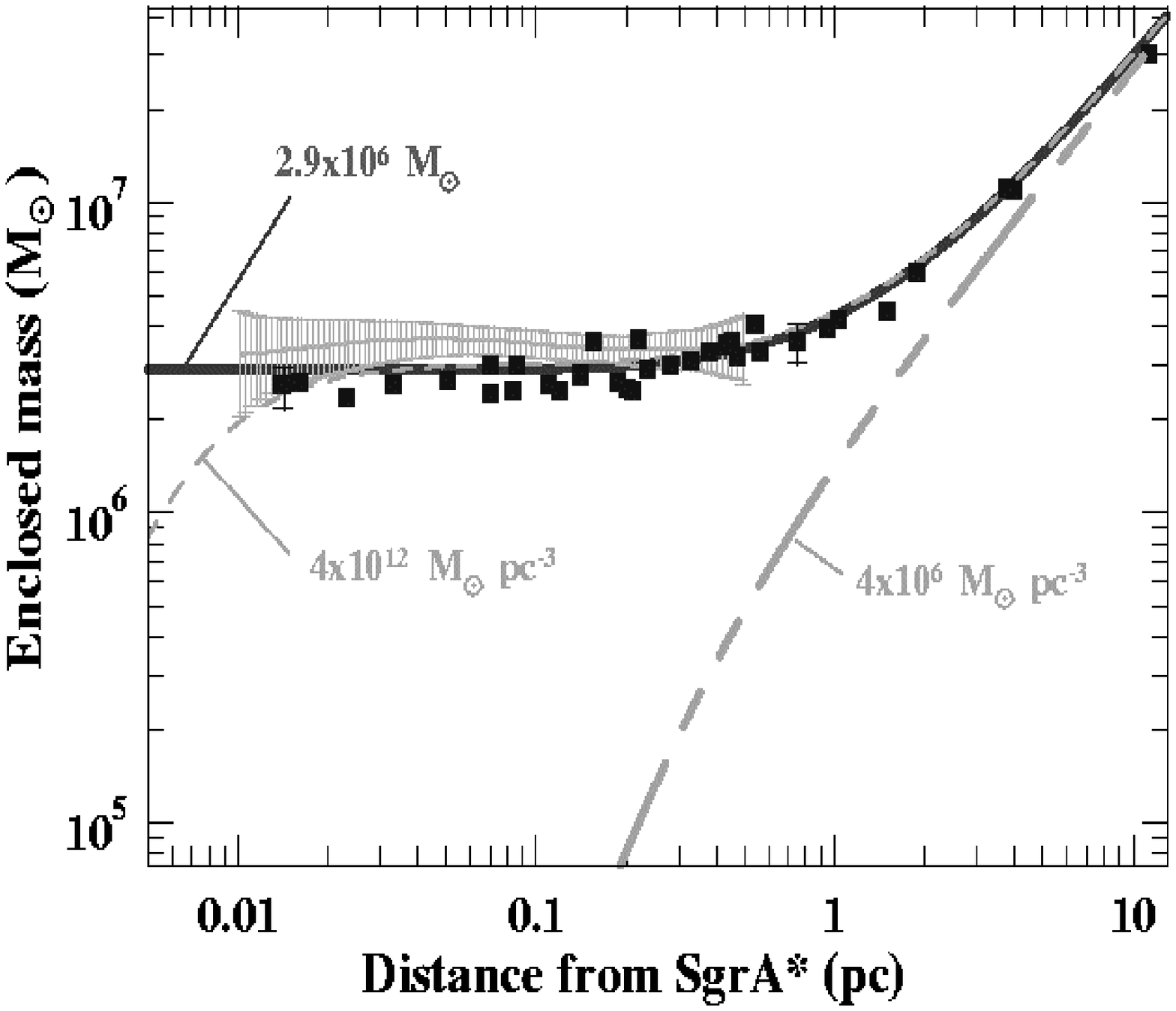}}
\end{document}